\documentclass[%
 reprint,
nofootinbib,
 amsmath,amssymb,
 aps,
]{revtex4-1}

\usepackage{graphicx}
\usepackage{bm}
\usepackage{hyperref}
\usepackage[mathlines]{lineno}

\usepackage[utf8]{inputenc}
\usepackage{amsmath,amssymb}
\usepackage[T1]{fontenc} 
\usepackage{microtype} 
\usepackage[english]{babel} 

\usepackage{booktabs} 

\usepackage{graphicx}
\usepackage{amssymb}
\usepackage{braket,mleftright}
\usepackage{empheq}
\usepackage{subfigure}
\usepackage{stackrel}
\usepackage{soul}
\usepackage{blkarray}
\usepackage{multirow}
\usepackage{amsmath}
\usepackage{physics}
\usepackage{amsfonts}
\usepackage{bm}
\usepackage{bbold} 
\usepackage{color}
\newcommand{\beq}{\begin{equation}}
\newcommand{\eeq}{\end{equation}}
\newcommand{\bse}{\begin{subequations}}
\newcommand{\ese}{\end{subequations}}
\newcommand{\bea}{\begin{eqnarray}}
\newcommand{\eea}{\end{eqnarray}}
\newcommand{\fabri}[1]{{\color{red} #1}}

\usepackage{enumitem} 
\setlist[itemize]{noitemsep} 

\usepackage{hyperref} 
\usepackage{braket}
\usepackage{verbatim} 

\begin{document}


\title{Quantum Speed Limit Bounds in an Open Quantum Evolution}

\author{Nicol\'{a}s Mirkin}
 \affiliation{
Departamento de F\'isica ``J. J. Giambiagi'' , FCEyN, Universidad de Buenos Aires, 1428 Buenos Aires, Argentina}
\author{Fabricio Toscano}
\affiliation{Instituto de F\'{\i}sica, Universidade Federal do Rio de
Janeiro, Caixa Postal 68528, Rio de Janeiro, RJ 21941-972, Brazil}

\author{Diego A. Wisniacki}
\affiliation{Departamento de F\'isica ``J. J. Giambiagi'' and IFIBA, FCEyN, Universidad de Buenos Aires, 1428 Buenos Aires, Argentina}





\begin{abstract}
Quantum mechanics dictates bounds for the minimal evolution time between 
predetermined initial and final states. Several of these
Quantum Speed Limit (QSL) bounds were derived for
non-unitary dynamics using different approaches. 
Here, we perform a systematic
analysis  of the most common QSL bounds 
in the damped Jaynes-Cummings model,
covering the Markovian and non-Markovian regime. 
We show that only one of the analysed bounds  
cleaves to the essence of the QSL theory outlined in the pioneer works
of Mandelstam \& Tamm and  Margolus \& Levitin in the context of unitary evolutions.
We also show that all of QSL bounds 
analysed reflect the fact  that in our model
non-Markovian effects speed up the quantum evolution. 
However, it is not possible to infer the  
 Markovian or non-Markovian behaviour of the dynamics 
 only analysing the QSL bounds.

\end{abstract}

\maketitle


\section{\label{Section-Intro}Introduction}

Knowing the fundamental limits that quantum mechanics imposes on the maximum speed 
of evolution between two distinguishable states is of utmost importance for quantum communication \cite{comm}, computation \cite{comp}, metrology \cite{metr} and many other areas of quantum physics.
In particular, the presence of decoherence \cite{deco,petruccione-book} makes the estimation of the minimal duration time of a process  of key  value in the designing of  quantum control protocols and in the implementation of quantum information tasks.
\par 
The Quantum Speed Limit (QSL) time, $\tau$, is defined as the minimal time a quantum system needs to evolve between an initial and a final state
separated by a given predetermined distance \cite{taddei,deffner}. The pioneering work on this subject was conducted by Mandelstam and Tamm (MT) \cite{MT}, who derived a bound for the evolution time of a system between two pure orthogonal states through a unitary dynamics generated by a time-independent Hamiltonian $\hat H$. 
 The resulting lower bound for the evolution time was given in $t \geq \tau^{MT}\equiv \hbar\pi/2\langle (\Delta \hat H)^2\rangle$ where
$\langle (\Delta \hat H)^2\rangle=\sqrt{\langle\hat H^{2} \rangle-\langle\hat H\rangle^{2}}$ denotes the variance of the energy of the system.
Several years later, Margolus and Levitin (ML) \cite{ML,levi}
studied the same problem and arrived to a different bound, i.e., $t \geq \tau^{ML} \equiv \pi \hbar /2\langle\hat H \rangle$, where $\langle\hat H \rangle$ is the mean energy. Therefore, for unitary dynamics connecting two orthogonal pure states, the bound for the quantum speed limit is not unique and the result was usually given by combining these two independent bounds and looking for the tightest: 
$t \geq max \left[\tau^{MT}; \tau^{ML}\right]$ \cite{Giovanetti2003}.
\par 
For non-unitary dynamics the extension of the MT approach was  given in 
\cite{taddei} using the Bures fidelity \cite{bures,Nielsen-book,mixed} between the initial and final states.
From their approach it can be extracted two bounds, that we call  $\tau_{t}^{min}$ and 
$\tau_{t}^{av}$.
The first minimal evolution time, $\tau_{t}^{min}$, corresponds to the time required by the process to traverse a distance equal
to the geodesic length between the two states, $\hat \rho_0$ and $\hat \rho_t$.
This time, can be estimated with few information about the dynamics and could depends on the actual time $t$, only implicitly 
through the state $\hat \rho_t$. 
\par
The second QSL bound, $\tau_{t}^{av}$
, involves a definition of an average speed of evolution, $\mathcal{V}^{av}_t$ (with frequency units), calculated in terms of the quantum Fisher information along the evolution path.
Both QSL bounds, $\tau_{t}^{min}$ and $\tau_{t}^{av}$, 
are tight for an evolution along the geodesic path between the initial and final states. 
This continuous in time tightness feature is important to engineering evolutions that achieve the minimal 
time of evolution set by quantum mechanics.
However, here we show that the explicit dependence of  the average velocity, $\mathcal{V}^{av}_t$, on the actual evolution 
time $t$, makes $\tau_{t}^{av}$ an inconsistent 
estimate of the minimal evolution time. This is shown in the well known damped Jaynes-Cummings (DJC) model.
On the contrary, $\tau_{t}^{min}$, gives a finite estimative of the minimal evolution time for all times for which the asymptotic state is essentially reached.
 \par
Other QSL bounds were given in literature for non-unitary evolutions \cite{delcampo,deffner,jing,Pires2016}.
Some of them \cite{deffner,jing} are also based on the definition
of velocities, $\mathcal{V}_t $ (with frequency units), that depend explicitly on the actual evolution time $t$. 
We show that all these bounds 
also give inconsistent estimates of the minimal evolution time. 
In the case of the QSL bound in \cite{deffner}, we have also demonstrated another drawback:
it does not own a continuously in time saturation, {\it i.e.} 
an evolution path where the bound is tight for all times.  
Thus, we argue that for non-unitary evolutions, $\tau^{min}_t$, is, within the analysed 
QSL bounds, the only one that sticks close to the essence of the QSL theory.
This essence is not to estimate the actual evolution time  but the 
minimal time needed to connect two states separated by a given distance.   
\par 
Other interesting aspect of the QSL bounds for open systems that was recently discussed in the literature is their connection with 
the non-Markovian character of the non-unitary evolutions  \cite{deffner, Sun2015,Meng2015}.
In fact, it was suggested in Ref. \cite{deffner} that one of their proposed QSL bounds could have enough information about the dynamics in order 
to be correlated with the Markovianity or non-Markovianity of the system evolution.
In particular, it was remarked that the non-Markovian effects, associated with the information back flow from the environment, could lead 
to faster quantum evolution and hence to smaller QSL times. 
Similar statements where made in \cite{Sun2015,Meng2015} and
we can say that it is widely spread \cite{Pires2016} the statement that: the
non-Markovianity speeds up the quantum evolution and that this feature can be infer from the behaviour of the QSL bounds.
Here, we consider the DJC model in the Rotated Wave Approximation (RWA) but with a detuning between the peak frequency of the spectral density and  the transition frequency of the qubit  whose dynamics can be tuned 
from essentially Markovian to a non-Markovian one.
We found that in the DJC model the non-Markovian effects  indeed speed-up the quantum evolution. 
Comparing 
all the QSL bounds analysed, in a wide range of parameters that controls the system,  with a measure of the non-Markovianity of the evolution,
we show that all of them are systematic smaller in the region of 
parameters corresponding to non-Markovian effects with respect
to their values in the region of parameters corresponding to a Markovian 
behaviour of the dynamics. In this sense we can say that 
the QSL bounds analysed reflects the speed-up  of the quantum evolution
due to non-Markovian effects in the DJC model.
However, we have shown that the converse it is not true, so there are regions of parameters that can not be associated with a non-Markovian behaviour of the dynamics where
the QSL bounds are as small as in the region of parameters where the dynamics 
is essentially non-Markovian. Therefore,
it is not possible to infer the speed up of the quantum evolution due to non-Markovian effects from the QSL bounds analyzed.
\par
The paper is organized as follows. In Section \ref{SectionII} we summarize the 
three different approaches for deriving the QSL bounds treated in this work, and analyse
the conditions for their saturation. Next, in Section \ref{SectionIII} we review the model used to test our statements: the DJC model  for zero temperature reservoir within the RWA, whose dynamics 
can be tuned from Markovian up to non-Markovian regimes. Our results are shown in Section \ref{Section-results}, and in Section \ref{Section-conclusion}, we conclude with some final remarks. 

\section{\label{SectionII}Quantum Speed Limit bounds for open systems}

A desirable feature for any QSL time bound is to be tight. 
This means that there is always an evolution that allows it saturation.  
Here, we summarize the derivation of the QSL bounds given in  \cite{taddei}, \cite{deffner} and \cite{jing} 
and we briefly analyse the conditions for their saturation. In particular, 
we focus on whether exist or not a continuous in time saturation, {\it i.e.}
a evolution path that for every time saturates the bound. 

\subsection{QSL bounds in terms of the quantum Fisher information.}
\label{SectionIIA}

The approach in \cite{taddei}  is based on the Bures fidelity \cite{Nielsen-book}
between the initial and final states, {\it i.e}
\beq
F_{B}(\hat \rho_0,\hat \rho_t)=Tr\left (\sqrt{\sqrt{\hat \rho_0}\hat \rho_t\sqrt{\hat \rho_0}}\right).
\eeq
The authors prove that, between all the metrics based on the Bures fidelity, the tightest lower bound for the Bures length \cite{Ulhmann-libro}, 
$\int_{0}^{t}\sqrt{\mathcal{F_{Q}}(t^\prime)/4}dt^\prime$,  is given by the Bures angle, 
$\arccos\left(F_{B}[\hat \rho_0,\hat \rho_t]\right)$ \cite{mixed,bures}, {\it i.e.}
\beq
\mathcal{L}(\hat \rho_0,\hat\rho_t)\equiv\arccos\left(F_{B}[\hat\rho_0,\hat\rho_t]\right)\leq \int_{0}^{\tau}\sqrt{\mathcal{F_{Q}}(t^\prime)/4}\;dt^\prime.
\label{taddei-ineq}
\eeq
Here, $\mathcal{F_{Q}}(t)$, is the quantum Fisher information along the path determined by the system evolution and its square root  is proportional
to the instantaneous speed of separation between two
neighbouring states. Eq. (\ref{taddei-ineq}) implies that the length of the geodesic 
that connects $\hat \rho(0)$ with $\hat \rho(t)$ is
always shorter than the length of the actual path.
\par
The geometric interpretation of Eq.(\ref{taddei-ineq}), 
allows to set up two types of minimal evolution 
time for two states separated by a given predetermined distance.
The first one, that we have called $\tau_{t}^{min}$, corresponds to the time the system it takes to
travel  (along the actual evolution path) the same length as the geodesic's length between the two states,
{\it i.e.}
\beq
\label{def-tau-min}
\mathcal{L}(\hat \rho_0,\hat\rho_t)= \int_{0}^{\tau^{min}_{t}}\sqrt{\mathcal{F_{Q}}(t^\prime)/4}\;dt^\prime.
\eeq 
It is important to realise that in order to know $\mathcal{F_{Q}}(t)$ along the path, in principle,
requires less information that to know exactly the actual dynamics of the system.
In this way, this QSL time follows the essence of the quantum speed limit theory because, knowing 
the initial and final state and without knowing the actual evolution time $t$, 
we can estimate a lower bound for the evolution time. This is well illustrated, 
for example, for any unitary evolution 
generated by a time-independent Hamiltonian, where $\mathcal{F}_Q(t)=4\langle (\Delta \hat H)^2\rangle_{\hat \rho_0}/\hbar^2$ for all times. So, in this case we only need 
the variance of the energy of the system  to estimate the bound,
\beq
\tau_{t}^{min}=\hbar \mathcal{L}(\hat \rho_0,\hat\rho_t)/
\langle (\Delta \hat H)^2\rangle_{\hat \rho_0},
\eeq
that for orthogonal pure states, {\it i.e.} $\mathcal{L}(\hat \rho_0,\hat\rho_t)=\pi/2$,  it is
equal to $\tau^{MT}$.
The QSL bound $\tau_{t}^{min}$ allows to define the speed limit ``velocity'' (with frequency units):
\beq
\label{def-vel-min}
\mathcal{V}^{min}_t\equiv\frac{\mathcal{L}(\hat \rho_0,\hat\rho_t)}{\tau_{t}^{min}},
\eeq
that depends on $t$ only implicitly through the final state $\hat \rho_t$.
\par

The second QSL bound comes directly from rearrenging  
Eq.(\ref{taddei-ineq}),
\beq
t\geq\frac{\mathcal{L}(\hat \rho_0,\hat\rho_t)}
{\mathcal{V}^{av}_t}   \equiv \tau_{t}^{av},
\label{QSL-time-GMT}
\eeq
where we define the ``average speed of the evolution'' as: 
 \beq
 \label{def-vel-GMT}
 \mathcal{V}^{av}_t \equiv(1/t)\int_{0}^{t}\sqrt{\mathcal{F_{Q}}(t^\prime)/4}\;dt^\prime.
 \eeq 
 In the case of unitary evolution generated by a time-independent Hamiltonian
we have  that
$ \mathcal{V}^{av}_t=\langle (\Delta \hat H)^2\rangle_{\hat \rho_0}/\hbar$,
does not depend on the actual time of evolution $t$,  and $\tau_{\tau}^{av}=\tau_{t}^{min}$.
 For non-unitary evolutions the times, $\tau_{t}^{av}$ 
and $\tau_{t}^{min}$, do not need to be equals, and 
in general, $\mathcal{V}^{av}_t$,
depends explicitly  on $t$, contrary to the velocity in $\mathcal{V}^{min}_t$  in Eq.(\ref{def-vel-min}).
Later on we will show, in a specific system, that
 $\tau_{t}^{min}<\tau_{t}^{av}$, and the explicit dependence 
 of $\mathcal{V}^{av}_t$ on $t$, makes $\tau_{t}^{av}$ an inconsistent 
 estimate of the minimal evolution time between 
 $\hat \rho_0$ and $\hat \rho_t$.
\par
 It is clear, from the geometric character of the inequality in Eq.(\ref{taddei-ineq}),
 that the saturation $\tau=\tau_{t}^{min}$ or $\tau =\tau_{t}^{av}$ is only possible
 whenever the system evolution is through a geodesic path, so in this case we have
 $\tau=\tau_{t}^{min}=\tau_{t}^{av}$ for all values of $t$. 
 Thus, both bounds,  $\tau=\tau_{t}^{min}$ and  $\tau=\tau_{t}^{av}$,
 are continuously tight, {\it i.e.} their saturation is
 continuously in the variable $t$ along the evolutions over geodesics.
 
\subsection{QSL bounds in terms of different operator norms. }
\label{SectionIIB}

Deffner and Lutz \cite{deffner} derived three different QSL bounds for a pure initial state $\rho_{0}=\ket{\psi_{0}}\bra{\psi_{0}}$ employing the von Neumann trace inequality for operators. 
Like in Ref. \cite{taddei} their approach also uses the Bures angle, $
\mathcal{L}(\rho_{0},\rho_{t})=\arccos(\sqrt{\expval{\hat \rho_t}{\psi_0}})$, 
in order to measure the predetermined distance
between the initial and final states.
The derivation can be summarized as follows.
First, from the time derivative of the Bures angle and using that $x\leq |x|$, it can arrive to
\begin{equation}
\begin{split}
\label{def-lutz-start-point}
    2\cos{(\mathcal{L})}\sin{(\mathcal{L})}\dot{\mathcal{L}} & \leq 
     |\expval{\dot{\hat \rho}_t}{\psi_0}|=
   \left |\Tr(\hat \rho_0\dot{\hat \rho}_t)\right|.
\end{split}
\end{equation}
 Next, it is used the von Neumann trace inequality for Hilbert-Schmidt class operators 
 \footnote{For two Schmidt class operators $\hat A$ and $\hat B$ the von Neumann trace inequality is
$\Tr(\hat A\hat B)\leq \sum_{i}\sigma_i\lambda_i$, where the sum is over the singular values, 
$\sigma_i$ and $\lambda_i$, of the operators, $\hat A$ and $\hat B$, respectively, in descending order, {\it} 
$\sigma_1\geq \sigma_2\geq \ldots$  and  $\lambda_1\geq \lambda_2\geq \ldots$ \cite{Grigorieff1991}.},
 \beq
 \label{ineq-trace}
  \left |\Tr(\hat \rho_0\dot{\hat \rho}_t)\right|\leq \sigma_1(t)= \|\dot{\hat \rho}_t\|_{op},
 \eeq
where $\sigma_1(t)$ is the largest singular value of $\dot{\hat \rho}_t$, 
and because this operator  is Hermitian, $\sigma_1(t)$ is equal to its operator norm
denoted by $\|\ldots\|_{op}$. 
Together with the inequality Eq.(\ref{ineq-trace}), it is used 
the set of inequalities for trace class operators,
\beq
\|\hat A\|_{op}\leq \|\hat A\|_{hs} \leq \|\hat A\|_{tr},
\eeq
where $\|A\|_{tr}\equiv\Tr(\sqrt{\hat A^\dagger\hat A})=\sum_i\sigma_i$ is the trace norm and $ \|\hat A\|_{hs}\equiv\sqrt{\Tr(\hat A^\dagger\hat A)}=\sqrt{\sum_i\sigma_i^2}$  is the Hilbert-Schmidt norm.
Gathering all the inequalities the authors arrive to,

\bea
2\cos{(\mathcal{L})}\sin{(\mathcal{L})}\dot{\mathcal{L}} \leq 
 \| \dot{\hat \rho}_t\|_{op}\leq \| \dot{\hat \rho}_t\|_{hs}
  \leq  \| \dot{\hat \rho}_t\|_{tr},
\eea
and integrating over time finally it is obtained,
\bea
 & \sin^2(\mathcal{L}(\rho_{0},\rho_{t})) \leq 
   \int_{0}^{t}\| \dot{\hat \rho}_{t^\prime}\|_{op}\;dt^\prime\leq \nonumber\\
   &\leq \int_{0}^{t}\;\| \dot{\hat \rho}_{t^\prime}\|_{hs}\;dt^\prime
   \leq  \int_{0}^{t}dt\;\| \dot{\hat \rho}_t\|_{tr}.
  \label{set-ine-deffner-lutz}
  \eea
\par
These inequalities are valid for any density operator evolution, 
and in the same way that Eq.(\ref{taddei-ineq}),
Eq.(\ref{set-ine-deffner-lutz}) serves as the starting point to derive QSL bounds if we define,
\beq
\label{def-vel-op}
\mathcal{V}_{t}^{op,tr,hs}\equiv (1/t)\int_{0}^{t}||L_{t}(\rho_{t^\prime})||_{op,tr,hs}\;dt^\prime.
\eeq
Then, the three QSL bounds derived in  \cite{deffner}  are:
\begin{equation}
\label{3-QSL-limits}
t \geq \tau_{t}^{op,tr,hs}=\frac{\sin^{2}{[\mathcal{L}(\rho_{0},\rho_{t})]}}{\mathcal{V}_{t}^{op,tr,hs}}.
\end{equation}
Because, $\mathcal{V}_{t}^{op}\leq \mathcal{V}_{t}^{hs}\leq \mathcal{V}_{t}^{tr}$, the greater QSL bound is $ \tau_{t}^{op}$.
Later on we will show, in a specific system, that
 $\tau_{t}^{op}>\tau_{t}^{min}$, and the explicit dependence 
 of $\mathcal{V}^{op}_t$ on $t$, makes of $\tau_{t}^{op}$ also an inconsistent 
 estimate of the minimal evolution time between 
 $\hat \rho_0$ and $\hat \rho_t$.
\par
We note that the inequalities in Eq.(\ref{set-ine-deffner-lutz}) have not 
a clear geometric interpretation, so the conditions for their saturation (that lead to the  
saturation of the QSL bounds in Eq.(\ref{3-QSL-limits})) are not so evident. 
In the case of the $\tau_{t}^{op}$, the saturation
corresponds to,
\beq
\label{satu-equality-tau-op}
\sin^2(\mathcal{L}(\rho_{0},\rho_{t})) = 
   \int_{0}^{t}dt\;\| \dot{\hat \rho}_t\|_{op}\l.
   \eeq
   \par
   In order to have a saturation over a given evolution path,
   we need to satisfy the equalities in Eq.(\ref{def-lutz-start-point}) and (\ref{ineq-trace}) for all times $t$. 
   So, the mean $\expval{\dot{\hat \rho}_t}{\psi_0}=\Tr(\hat \rho_0\dot{\hat \rho}_t)$ should be positive along the path.
   Let's suppose that this is the case so now we want to see 
   if it is possible to saturate Eq.(\ref{ineq-trace}) for all times $t$, {\it i.e.} 
    $ \Tr(\hat \rho_0\dot{\hat \rho}_t)= \sigma_1(t)\equiv \|\dot{\hat \rho}_t\|_{op}>0$
    along some evolution path. 
    In order to see that this is not possible, we first observe that the von Neumann trace inequality
    $ \Tr(\hat \rho_0\dot{\hat \rho}_t)\leq \sigma_1(t)$
    is saturated along an evolution path iff $\hat \rho_0$ and $\dot{\hat \rho}_t$ are 
   simultaneously unitarily diagonalisable for all evolution times. This means 
   that $\sigma_1(t)$ must be the eigenvalue of $\dot {\hat \rho}_t$
   associated with the time independent common eigenvector, $\ket{\psi_0}$, of
  $\dot {\hat \rho}_t$ and $\hat \rho_0$. Therefore, the structure of the evolved state
  should be
  \beq
  \label{evol-state-non-physical}
  \hat \rho_t=\left(1+\int_0^t \, \sigma_1(t^\prime)\,dt^\prime\right)\hat \rho_0+
  \hat A_t,
  \eeq
   where $\hat A_t$ has a support in the subspace orthogonal to 
   the subspace spanned by $\hat \rho_0\equiv \dyad{\psi_0}{\psi_0}$. 
   But because we assume that Eq.(\ref{def-lutz-start-point}) is saturated for all times,
   we have 
   $\sigma_1(t)>0$ for all times. So, $\hat \rho_t$ in Eq.(\ref{evol-state-non-physical})
   is not a physical state for all $t>0$, because otherwise we would have for the probability 
   to find the evolved state in the initial state:
   \beq
   Tr(\hat \rho_0 \hat \rho_t)=1+\int_0^t \, \sigma_1(t^\prime)\,dt^\prime>1,
   \eeq
   where we use that $\hat \rho_0 \hat A_t=0$ for all times.
   Therefore it is not possible to find an evolution path where  Eq.(\ref{ineq-trace}) is saturated for all times if Eq.(\ref{def-lutz-start-point})
   is also saturated for all times. 
   The  saturation $t=\tau_{t}^{op}$ , can only be possible for certain times 
   $t$ 
   along a given path of the system evolution.
     This contrasts clearly to $t = \tau_{t}^{av}=\tau_{t}^{min}$, that is a continuously in time saturation along 
a geodesic evolution path. 
 \par  
\subsection{QSL bound using the notion of Quantumness.}
\label{SectionIIC}
\indent
The derivation of a QSL bound in \cite{jing} follows a very different approach
 based on the notion of ``quantumness''.
The quantification of the non-classical character of a quantum system has recently attracted 
much attention  \cite{iyengar,ferro}.   
In particular it was defined the notion of quantumness associated with  the non-commutativity of the algebra of observables \cite{iyengar,ferro} as, 
\begin{equation}
\begin{split}
Q(\hat \rho_{a},\hat \rho_{b}) & =2\|[\hat \rho_{a},\hat \rho_{b}]\|_{hs}^{2} \\
& =-4Tr\left [(\hat\rho_{a}\hat\rho_{b})^{2}-\hat \rho_{a}^{2}\hat \rho_{a}^{2} \right ],
\end{split}
\label{quantum}
\end{equation}
such $0\leq Q(\hat \rho_{a}, \hat \rho_{b}) \leq 1$. 
Note, that $Q(\hat \rho_{a}, \hat \rho_{b}) = 0$ iff $[\hat \rho_{a}, \hat\rho_{b}] = 0$ \cite{iyengar,ferro},
that it means that $\hat \rho_{a}$ and $\hat \rho_{b}$ are diagonal in the same basis.
In that sense $Q(\hat \rho_{a}, \hat \rho_{b}) $ is a witness of the coherences
that the state $\hat \rho_{b}$ has in the basis of eigenstates of $\hat \rho_a$ and vice versa. 
Therefore, in a system evolution, the quantumness, $Q(\hat \rho_{0}, \hat \rho_{t})$, as a function of time, monitors the generation of coherences in the evolved state 
$\hat \rho_t$, in the eigenstates basis of the initial state $\hat \rho_0$. 
\par
Contrary to the approaches described in the previous sections, in order to get a QSL bound,
the approach in  \cite{jing}  does not use explicitly any distance 
between the initial and final state. Instead, from
the definition of the quantumness  $Q(\hat \rho_{0}, \hat \rho_{t})$, the authors use the Cauchy-Schwarz inequality, 
{\it i.e.} $|\Tr(\hat A^\dagger \hat B )|\leq \|\hat A\|_{hs}\|\hat B\|_{hs}$, to obtain
\beq
\frac{|\dot{Q}(\hat \rho_{0}, \hat \rho_{t})|}{\sqrt{Q(\hat \rho_{0}, \hat \rho_{t})}}\leq 2\sqrt{2}\;\|[\hat \rho_0,\dot{\hat \rho}_t]\|_{hs},
\label{1st-del-campo-ineq}
\eeq
where $\dot{Q}(\hat \rho_{0}, \hat \rho_{t})=4\Tr(\hat A_t^\dagger\hat B_t)$ with
$\hat A_t\equiv [\hat \rho_0,\hat \rho_t]$ and $\hat B_t\equiv[\hat \rho_0,\dot{\hat \rho}_t]$.
Now, for the integration in time of the l.h.s. in Eq.(\ref{1st-del-campo-ineq}),
it is used that 
\beq
\int_{0}^t \;\frac{|\dot{Q}(\hat \rho_{0}, \hat \rho_{t^\prime})|}{\sqrt{Q(\hat \rho_{0}, \hat \rho_{t^\prime})}}\;dt^\prime\geq \left|\int_{0}^Q\frac{dQ^\prime}{\sqrt{Q^\prime}}\right|=
2\sqrt{Q(\hat \rho_{0}, \hat \rho_{t})}.
\label{2nd-del-campo-ineq}
\eeq
Therefore, they finally obtain,
\beq
\label{main-ineq-del-campo}
\sqrt{Q(\hat \rho_{0}, \hat \rho_{t})/2} \leq \int_{0}^{t} \;\|[\hat \rho_0,\dot{\hat \rho}_{t^\prime}]\|_{hs}
 \;dt^\prime.
\eeq
A QSL bound, $\tau_{t}^{quant}$, can be set up from the inequality in Eq.(\ref{main-ineq-del-campo}),
in the same way that, $\tau_{t}^{av}$, was set up  from the inequality in Eqs.(\ref{taddei-ineq})
or the bounds, $\tau_{t}^{op,tr,hs}$, from the inequalities in Eq.(\ref{set-ine-deffner-lutz}), {\it i.e.},
\begin{equation}
t\geq\tau_{t}^{quant}=
\frac{\sqrt{Q(\hat \rho_{0},\hat \rho_{t})/2}}{\mathcal{V}^{quant}_t},
\label{quantec}
\end{equation}
where we define the time average velocity with frequency units,
\beq
\label{def-vel-quant}
\mathcal{V}^{quant}_t\equiv \frac{1}{t} \int_{0}^t  \;\|[\hat \rho_0,\dot{\hat \rho}_{t^\prime}]\|_{hs}\;dt^\prime.
\eeq
\par 
In order to have a saturation in Eq.(\ref{main-ineq-del-campo}), therefore $t=\tau_{t}^{quant}$, for all times over a given evolution path,
   we need to satisfy the equalities in Eq.(\ref{1st-del-campo-ineq}) and (\ref{2nd-del-campo-ineq}) for all times $t$. 
Let's suppose that the rate of change of the quantumness, $\dot{Q}(\hat \rho_{0}, \hat \rho_{t})$, is positive along the evolution path, so the equality Eq.(\ref{2nd-del-campo-ineq}) is saturated along the path. 
This means that the rate of generation of coherences in $\hat \rho_t$, in the basis of eigenstates of $\hat \rho_0$, is positive for all times; something that could be possible.
In order  to saturate Eq.(\ref{1st-del-campo-ineq})
for all times along some evolution path, we need that,
\beq
\hat B_t=
\xi_t\hat A_t,
\eeq
with $\xi_t$ a real function of time.  Because we assume $\dot{Q}(\hat \rho_{0}, \hat \rho_{t})=4\xi_t \Tr(\hat A_t^\dagger\hat A_t)\ge 0$ we have that $\xi_t
\ge 0$ for all times.
This means that: {\it i)} $\dot{\hat \rho}_t=\xi_t \hat \rho_t$ or that {\it ii)} $\hat \rho_0$ and $\dot{\hat \rho}_t-\xi_t \hat \rho_t$ are
diagonal in the same basis,  for all times along some evolution path.
The option {\it i) } it is not possible because imposing the normalisation condition 
on the evolved state we arrive to $\int^t_0\;\xi_{t^\prime}\,dt^\prime=0$ for all times, condition
that can not be satisfy unless $\xi_t=0$ for all times. But, $\xi_t=0$ for all times,
corresponds to the trivial evolution where the evolved state remains equal 
to $\hat \rho_0$ for all times. However, 
the condition {\it ii)} can be satisfied for example in the cases of quasi-classical models
consisting of evolved states diagonal in the eigenbasis of the initial state 
$\hat \rho_0$ for all times, with 
only their eigenvalues changing along the evolution path \cite{Sorovar2006}.
Therefore, the QSL bound, $\tau_t^{av}$, in principle, can be saturated continuously in 
time along some evolutions paths.
\par
\section{\label{SectionIII}The Jaynes-Cummings model for zero temperature
reservoir}

\par
\renewcommand{\figurename}{Figure} 
\begin{figure}[!htb]
\begin{center}
\includegraphics[scale=0.60]{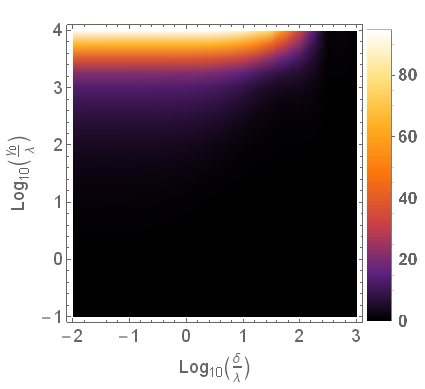}
\begin{footnotesize}
\caption{
 Density plot  of the non-Markovianity of the channel corresponding to the 
DJC model, measured by the expression in Eq.(\ref{breuer-measure}) for
the initial states,  $\hat\rho_{e}=\dyad{x;+}{x;+}$ and $\hat\rho_{g}=\dyad{x;-}{x;-}$, and a total time of evolution 
such that $\lambda t=1000$. See main text for details.}
\label{fig1}
\end{footnotesize}
\end{center}
\end{figure}
\par
In this section, we present a simple physical model that will serve as a platform to study all the QSL bounds presented in the previous section. We consider the exactly solvable damped Jaynes-Cummings model for a two-level system interacting
with a bosonic quantum reservoir at zero temperature, both in the resonant and the detuning regime \cite{breuer43,garraway,petruccione-book,breuer2,He2011,pineda,
Bylicka2014}. 
The Hamiltonian of the system is given by $\hat H=\hat H_{0}+\hat H_{I}$. 
The free Hamiltonian of the qubit and the modes of the reservoir is, $\hat H_{0}=\omega_{0}\hat \sigma_{z}+\sum_{k}\omega_{k}\hat b_{k}^{\dagger}\hat b_{k}$,  while, $\hat H_{I}=
\sum_k g_k\hat b_k\hat \sigma_{+}+g^*_k\hat b^\dagger_k\hat \sigma_{-}$, is the interaction Hamiltonian between them ($g_{k}$ is the coupling strength between the 
qubit and the mode $k$). 
Here, $\omega_{0}$, is the energy difference between the two levels system, $\hat \sigma_{\pm}$ are the rising and lowering operators for the qubit and 
$\hat \sigma_{z}$ is a Pauli operator.
The operators, $\hat b_{k}^{\dagger}$ and $\hat b_{k}$, are the creation and annihilation operators for the bosonic modes whose frequencies are $\omega_{k}$. 
In the limit of an infinite number of reservoir modes and a smooth spectral density, this model leads to the reduce qubit's evolution given by the exact master equation,
\bea
\label{master-equation}
&\dot{\hat \rho}_t=-\frac{is_t}{2}[\hat \sigma_z,\hat \rho_t]+\nonumber\\
&+\gamma_{t}\left(\hat \sigma_{-}\hat \rho_{t}\hat \sigma_{+}-\frac{1}{2}\hat\sigma_{+}\hat \sigma_{-}\hat \rho_{t}-\frac{1}{2}\hat \rho_{t}\hat \sigma_{+}\hat \sigma_{-}  \right),
\eea
with $s_t=-2\mathbb{I}m\{\dot{G}(t)/G(t)\}$ and $\gamma_t=-2\mathbb{R}e\{\dot{G}(t)/G(t)\}$
the time-dependent Lamb shift and the decay rate respectively \cite{petruccione-book}. 
The solution of this master equation is given by the channel \cite{He2011,petruccione-book}:
\begin{equation}
\label{JC-evolution}
\hat \rho_t=\Lambda_{t}[\hat \rho_0]=\left[
  \begin{array}{ c c }
     |G(t)|^{2}\rho_{ee} & G(t)\rho_{eg} \\
     G(t)^{*}\rho_{eg}^{*} &1- |G(t)|^{2}\rho_{ee}
  \end{array} \right],
\end{equation}
where the initial state of the qubit is 
\beq
\hat \rho_0=\left[
  \begin{array}{ c c }
     \rho_{ee} & \rho_{eg} \\
     \rho_{eg}^{*} & 1-\rho_{ee}
  \end{array} \right],
  \eeq 
  in the basis, $\ket{z;\pm}$, 
of eigenstates of the free Hamiltonian of the qubit.
  The function, $G(t)$, is the solution to the equation 
  $\dot{G}(t)=-\int_{0}^{t}d\tau f(t-\tau)G(\tau)$, with $G(0)=1$ and where $f(t-\tau)$ is the two point correlation function of the reservoir, {\it i.e.} the Fourier transform of the spectral density $J(\omega)$. 
For a Lorenzian spectral density, $J(\omega)=\gamma_0\lambda^2/2\pi[(\omega-\omega_c)^2+\lambda^2]$, ($\lambda$ is its width, $\omega_c$ is its peak frequency and $\gamma_0$ is an effective coupling constant) it is obtained the result \cite{pineda},
\begin{equation}
f(t)=\frac{1}{2}\gamma_{0}\lambda e^{-\lambda|t|(1-i \delta/\lambda)},
\end{equation}
with $\delta=\omega_0-\omega_c$  the detuning between the peak frequency of the 
spectral density and the transition frequency of the qubit.
Therefore, 
\bea
G(t)&=&e^{-\frac{\lambda\,t}{2}\left(1-i\frac{\delta}{\lambda}\right)}
\left[\frac{\lambda}{\Omega}\left(1-i\frac{\delta}{\lambda}\right)\sinh\left(\frac{\Omega t}{2} \right)+\right.\nonumber\\
&&+\left.\cosh\left(\frac{\Omega t}{2} \right) \right],
\eea
where $\Omega=\lambda\sqrt{(1-i \delta/\lambda)^{2}-2\gamma_{0} /\lambda}$ and  
the time-dependent decay rate is,
\begin{equation}
\gamma_{t}=\gamma_{0}\;\mathbb{R}e\left(\frac{ \;2
\sinh\left(\frac{\Omega t}{2}\right)}{\frac{\Omega}{\lambda}   \cosh\left(\frac{\Omega t}{2}\right)+(1 -i\frac{\delta}{\lambda})  \sinh\left(\frac{\Omega t}{2}\right)}\right).
\end{equation}
Note, that if we measure the time in units of $1/\lambda$, the function $G(t)$ and therefore the decay rate $\gamma_t$, depends only on two parameters, {\it i.e.} $\gamma_0/\lambda$ and 
$\delta/\lambda$.
\par
An important feature of the DJC model is that have different regimes of the parameters, 
$\gamma_0/\lambda$ and $\delta/\lambda$, that can be associated with Markovian and non-Markovian effects on the evolution. 
In the limit, $\gamma_0/\lambda\ll 1$ and 
$\delta/\lambda\ll 1$, we get for the decay rate: $\gamma_t=\gamma_0/(1+\coth(\lambda t/2))$, which it is  
a strictly increasing positive function of time, that when $\lambda t \gg 1$ it corresponds to $\gamma_t\sim \gamma_0$.
Because of $\gamma_t=\gamma_0/(1+\coth(\lambda t/2))\ge 0$ for all times,
Eq.(\ref{master-equation}) is a Markovian master equation \cite{breuer2}
in the regime $\gamma_0/\lambda\ll 1$ and 
$\delta/\lambda\ll 1$. However, away from this regime, 
in order to check Markovianity or non-Markovianity of the dynamics, it is necessary to monitor the distinguishability 
between any two states along the evolution.
This is because the accepted notion of Markovianity that we will use here is based on the idea that for Markovian processes any two quantum states become less and less distinguishable under the dynamics, leading to a continuous loss of information into the environment  \cite{breuer2}. 
\par
The trace norm of the difference, $\hat \rho_{1}-\hat \rho_{2}$, it is used to define the trace distance,
\begin{equation}
   D(\hat \rho_{1},\hat \rho_{2})=\frac{1}{2}||\hat \rho_{1}-\hat \rho_{2}||_{tr}=\frac{1}{2}\Tr|\hat \rho_{1}-\hat \rho_{2}|,
\end{equation}
that is a measure of the distance between the two quantum states \cite{Nielsen-book}. 
This measure has the nice property that can be interpreted as a measure of 
distinguishability between $\hat \rho_1$ and $\hat \rho_2$ \cite{breuer}.
Therefore, based on the trace distance, the characterization of the non-Markovian 
character of a quantum process, given by the map $\hat \rho_t=\Lambda_{t}[\hat \rho_0]$, 
can be stated as: 
a quantum map $\hat \rho_t=\Lambda_{t}[\hat \rho_0]$ is non-Markovian
if and only if there is a pair of initial states, $\hat\rho_{0,1}$ and $\hat \rho_{0,2}$, such that the trace distance between the corresponding evolved states increases at a certain time $t$, i.e.
\begin{equation}
    \sigma(t, \hat\rho_{0,1},\hat \rho_{0,2})\equiv
    \frac{d}{dt}D(\Lambda_{t}[\hat \rho_{0,1}],\Lambda_{t}[\hat \rho_{0,2}])>0,
\end{equation}
where $\sigma(t, \hat\rho_{0,1},\hat \rho_{0,2})$ denotes the rate of change of the trace distance at time $t$ corresponding to the initial pair of states \cite{breuer}. 
For a non-Markovian process, information must flow from the environment to the system for some interval of time, and thus we must have $\sigma>0$ for this time interval. A good measure of non-Markovianity of the channel should witness the total increase of the distinguishability over the whole time evolution, i.e. the total amount of information flowing from the environment back to the system. This suggests defining a measure $\mathcal{N}(\Lambda_t)$ for the non-Markovianity of a quantum process through \cite{breuer2}:
\beq
\mathcal{N}(\Lambda_t)=max_{\{\hat\rho_{0,1},\hat \rho_{0,2}\}} \;\mathcal{N}(\Lambda_t,\hat\rho_{0,1},\hat \rho_{0,2}),
\label{breuer-measure}
\eeq
with
\beq
\mathcal{N}(\Lambda_t,\hat\rho_{0,1},\hat \rho_{0,2})=\int_{\sigma>0}\,dt\;
\sigma(t,\hat\rho_{0,1},\hat \rho_{0,2}).
\eeq
For a general process the maximisation over the initial states, $\hat \rho_{0,1}$ and $\hat \rho_{0,2}$, in $\mathcal{N}(\Lambda_t)$, is a difficult task.
However for the DJC model considered here, when $\delta\neq 0$,  it was shown in \cite{breuer} that
$\mathcal{N}(\Lambda_t)=\mathcal{N}(\Lambda_t,\hat\rho_{e},\hat \rho_{g})$
where $\hat\rho_{e}=\dyad{x;+}{x;+}$ and $\hat\rho_{g}=\dyad{x;-}{x;-}$,
with $\ket{x;\pm}$ the eigenstates of the Pauli operator $\hat \sigma_x$.
In Fig.\ref{fig1} we show the behaviour of the measure $\mathcal{N}(\Lambda_t)$
as a function of the parameters $\gamma_0/\lambda$ and $\delta/\lambda$ that control the DJC model.
\section{\label{Section-results}Results: QSL bounds in the DJC model}

\begin{figure}[!htb]
\includegraphics[scale=0.8]{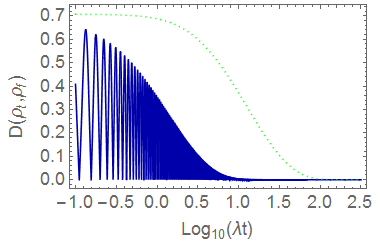}
\caption{(Color online)Trace distance $D(\hat \rho_t,\hat \rho_f)$ between the final stationary state  $\hat \rho_f=\dyad{z;-}{z;-}$ and the evolved state $\hat \rho_t$ of the qubit in the DJC model, as a funtion 
of scaled time $\lambda t$. 
Green dotted line corresponds to the Markovian regime, 
with $\delta/\lambda=0.1$ and $\gamma_0/\lambda=0.1$, and the 
 the blue full line to the non-Markovian regime with 
 $\delta/\lambda=0.1$ and $\gamma_0/\lambda=10^4$.
 In both cases the initial state of the evolution is  $\hat \rho_i=\dyad{x;+}{x;+}$.}
\label{fig2}
\end{figure}
\par

The DJC model is a very suitable framework  to analyse all the QSL bounds discussed in the previous Section.
Our goal is to examine which of the bounds stay close to the essence of the QSL 
theory giving  consistent estimates for the minimal evolution time
to reach a final state from an  initial one within the framework of open quantum evolutions.
\par
The reduced evolution of the qubit in the DJC model in Eq.(\ref{JC-evolution})
has a stationary state, for all values of the parameters $\delta/\lambda$ and $\gamma_0/\lambda$. 
Indeed, no matter which is the initial state and due to the fact that 
$\lim_{t\rightarrow \infty}G(t)=0$,  the asymptotic final state is  $\hat \rho_f=\dyad{z;-}{z;-}$. The speed at which an evolved state approaches the stationary
state is different in the Markovian and non-Markovian regimes.
This is clearly shown in  Fig.\ref{fig2} where we plot the trace distance, 
$D(\hat \rho_t,\hat \rho_f)$, 
between the evolved state of the qubit $\hat \rho_t $ and its stationary state 
$\hat \rho_f$,
as a function of time for two different parameters that controls the environment and its interaction with the qubit. 
The initial state is $\hat \rho_i=\dyad{x;+}{x;+}$, however similar results were obtained 
from any other $\hat \rho_i$ (not shown).
We see that in the Markovian regime ($\delta/\lambda=0.1$ and $\gamma_0/\lambda=0.1$) the stationary state is reached for times $\lambda t > 100$ , while 
in the non-Markovian regime ($\delta/\lambda=0.1$ and $\gamma_0/\lambda=10000$) the final state $\hat \rho_f$ is approached for earlier times ($\lambda t  \approx 16)$. This shows the speed up of the evolution in the non-Markovian regime.
 
\begin{figure}[!htb]
\includegraphics[scale=0.8]{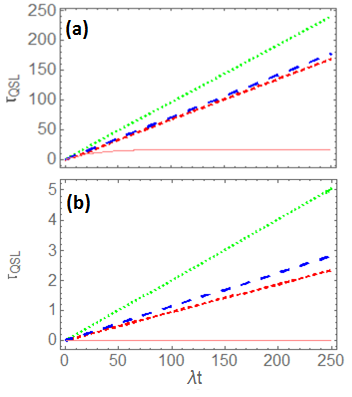}
\caption{(Color online) QSL bounds as a function of the final time of the evolution
corresponding to the DJC model, starting with the initial state of the qubit
$\hat \rho_i=\dyad{x;+}{x;+}$. In {\bf (a)} we show an example in the Markovian regime
with $\delta/\lambda=0.1$ and $\gamma_0/\lambda=0.1$, and {\bf (b)}
in the non-Markovian regime with 
 $\delta/\lambda=0.1$ and $\gamma_0/\lambda=10^4$.
Green dotted line is for $\tau_{t}^{quant}$, red dashed lines is for $\tau_{t}^{op}$, blue large-dashed line is for $\tau_{t}^{av}$ 
 and red solid line is for $\tau_{\tau}^{min}$.} 
\label{fig3}
\end{figure}
\par
\begin{figure}[!htb]
\includegraphics[scale=0.8]{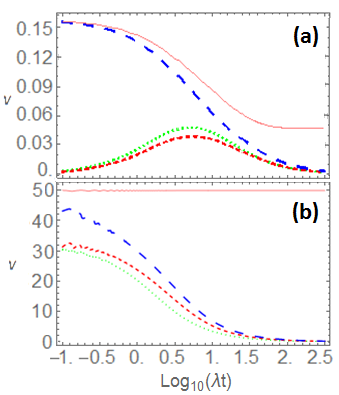}
\caption{ (Color online)  The average speed  as a function of the final time of the evolution
in the DJC model, with the initial state of the qubit
$\hat \rho_i=\dyad{x;+}{x;+}$.  In {\bf (a)} we show an example in the Markovian regime
with $\delta/\lambda=0.1$ and $\gamma_0/\lambda=0.1$, and {\bf (b)}
in the non-Markovian regime with 
 $\delta/\lambda=0.1$ and $\gamma_0/\lambda=10^4$.
 Green dotted line is for $\mathcal{V}_{t}^{quant}$ (Eq.(\ref{def-vel-quant})), red dashed line is for $\mathcal{V}_{t}^{op}$ (Eq.(\ref{def-vel-op})), blue large-dashed line is for $\mathcal{V}_{t}^{av}$  (Eq.(\ref{QSL-time-GMT})) and 
 red solid is for $\mathcal{V}^{min}_t$ (Eq.(\ref{def-vel-min})).}
\label{fig4}
\end{figure}

Let us now consider the behaviour  of the different QSL bounds as a function of the final 
time of evolution $\lambda t$ shown 
in  Fig.\ref{fig3}. We remark that 
equivalent results were obtained for any other initial  pure state (not shown).
We can appreciate in Fig.\ref{fig3} that for times  $\lambda t>100$
when, either in the Markovian and non-Markovian regime, the qubit  have reached
the stationary state $\hat \rho_f$ (see Fig.(\ref{fig2})),  only the bound $\tau_{t}^{min}$ remains constant. The other bounds grow approximately linear. 
This behavior  is due to the fact that in the denominator of the definitions of 
$\tau_{t}^{av}$, $\tau_{t}^{op}$ and $\tau_{t}^{quant}$ (Eqs.(\ref{QSL-time-GMT}), (\ref{3-QSL-limits}) and (\ref{quantec}) respectively),
appear the ``average velocities" (with frequency units), $\mathcal{V}_{t}^{av}$, $\mathcal{V}_{t}^{op}$ and $\mathcal{V}_{t}^{quant}$, that depends on the actual evolution time $t$.  These average velocities go to zero when the stationary state is achieved while the quantities in the
numerator of the definitions of the bounds remain constant. 
This is shown in Fig.\ref{fig4} where we plot  $\mathcal{V}_{t}^{av}$, $\mathcal{V}_{t}^{op}$ and $\mathcal{V}_{t}^{quant}$, as a function of the evolution time $\lambda t$, and we also plot $\mathcal{V}_{t}^{min}$ that was defined in Eq. \ref{def-vel-min}. 

The results shown in Fig.\ref{fig3}  and Fig.\ref{fig4} clearly  show that none of the bounds, $\tau_{t}^{av}$, $\tau_{t}^{op}$ and $\tau_{t}^{quant}$, give a consistent estimate of the 
minimal time to achieve the final state $\hat \rho_f=\dyad{z;-}{z;-}$ starting from the initial one 
$\hat \rho_i=\dyad{x;+}{x;+}$. 
Moreover, the average velocities, $\mathcal{V}_{t}^{av}$, $\mathcal{V}_{t}^{op}$ and $\mathcal{V}_{t}^{quant}$, have the same asymptotic behavior as the instant speed of evolution, given by  $\sqrt{\mathcal{F}_Q(\tau)/4}$, that for
$\lambda \tau > 100$ also it goes to zero.  
This fact goes against the essence of the 
QSL theory that pursue the estimation  of a speed limit velocity of the evolution 
between two states.
On the contrary, $\tau_{t}^{min}$ gives
a consistent estimate of the minimal time needed to reach $\hat \rho_f$
from $\hat \rho_i$, and also provides a quantum speed limit of evolution.
 
Although we have shown that only  one of the  QSL bounds presented in Section \ref{SectionII}  gives 
a reliable estimate of the minimum evolution time, we  study now the connection of these bounds with the non-Markovianity 
character of the evolution
 \cite{deffner, Sun2015,Meng2015}.
 
The measure $\mathcal{N}(\Lambda_t)$ in Eq.(\ref{breuer-measure}) is suitable to 
characterize the degree of non-Markovianity of a quantum channel $\Lambda_t$.
However, in order to establish a possible link between the QSL bounds 
and non-Markovian effects of the dynamics it is more appropriate to define a measure 
of non-Markovianity over the actual trajectory of the system {\it i.e.} 
from the initial state $\hat \rho_0$ to the final one $\hat \rho_t$, that enters in the definition of the QSL bounds.
In this way, we define
\bea
\tilde{\mathcal{N}}(t ;\Lambda_t,\hat\rho_{0})&=&
\int_{0,\tilde{\sigma}>0}^t\,
\tilde{\sigma}(t^\prime,\hat\rho_{0},\hat \rho_{t^\prime})\;dt^\prime=\nonumber\\
&=&
\int_{0}^t\,
\frac{|\tilde{\sigma}(t^\prime,\hat\rho_{0},\hat \rho_{t^\prime})|+\tilde{\sigma}(t^\prime,\hat\rho_{0},\hat \rho_{t^\prime})}{2}\;dt^\prime,
\label{our-measure}
\eea
that depends on the final time $t$, and where 
\beq
\tilde{\sigma}(t, \hat\rho_{0},\hat \rho_{t})\equiv
    \frac{d}{dt}D(\hat \rho_{0},\Lambda_{t}[\hat \rho_{0}]).
    \eeq

\begin{figure}[!htb]
\begin{center}
\includegraphics[scale=0.50]{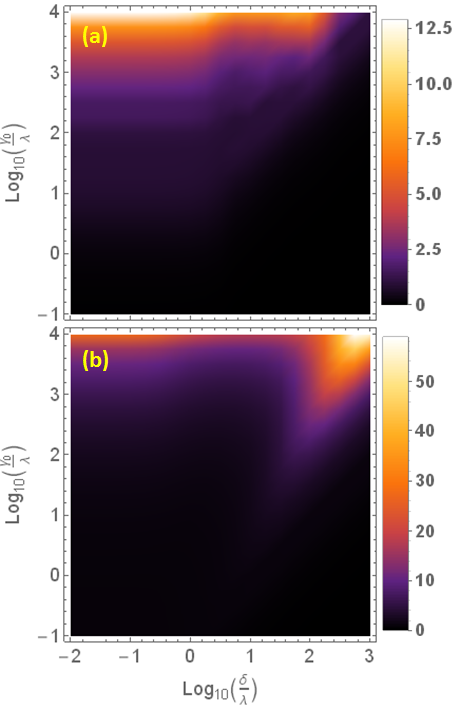}
\begin{footnotesize}
\caption{(Color online) 
Density plot of the non-Markovianity over a state evolution path in the 
DJC model, measured by the expression in Eq.(\ref{our-measure}),
calculated from the initial state,  $\hat\rho_{e}=\dyad{x;+}{x;+}$.  The time of evolution is 
$\lambda t=1$ in {\bf (a)} and $\lambda t=100$ in {\bf (b)}.}
\label{fig5}
\end{footnotesize}
\end{center}
\end{figure}

In Fig.(\ref{fig5}) we show a density plot of $\tilde{\mathcal{N}}(t ;\Lambda_t,\hat\rho_{0})$ as a function of the parameters $\gamma_0/\lambda$ and $\delta/\lambda$ for an initial state $\hat \rho_i=\dyad{x;+}{x;+}$ and two final evolution times: $\lambda t=1$ (panel {\bf (a)})
and $\lambda t=100$ (panel {\bf (b)}). Comparing  Fig.(\ref{fig1}) and  Fig.(\ref{fig5}), we can see similar qualitative behaviour of the two measures of the non-Markovianity as a function of the two parameters, $\gamma_0/\lambda$ and $\delta/\lambda$, that controls the dynamics of the channel.

\begin{figure*}[!htb]
\includegraphics[scale=0.45]{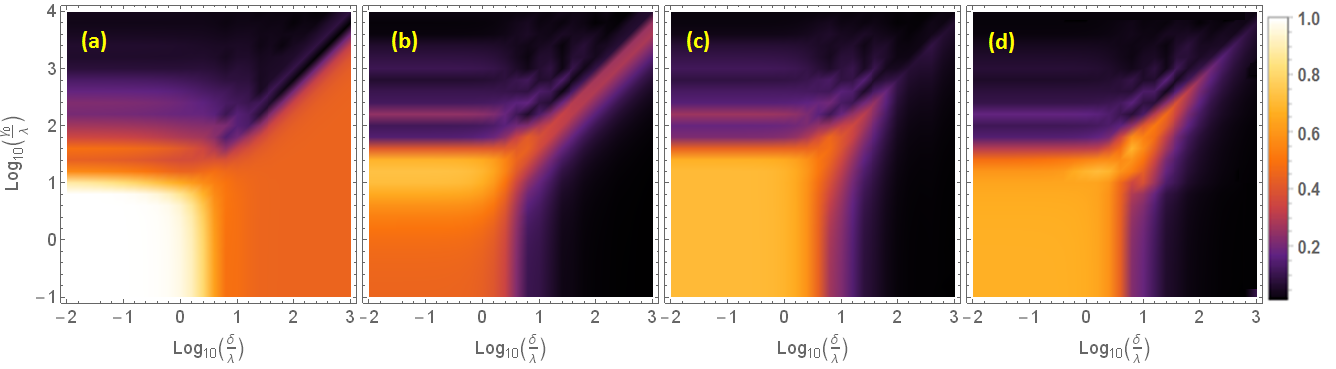}
\caption{(Color online)  Density plot of the QSL bounds as a function of the  parameters $\gamma_0/\lambda$ and $\delta/\lambda$.
The initial state is $\hat\rho_{i}=\dyad{x;+}{x;+}$ and the final time of evolution is $\lambda t=1$. In {\bf (a)} we plot  $\tau_{t}^{quant}$,
in  {\bf (b)} $\tau_{t}^{op}$, in {\bf (c)} $\tau_{t}^{av}$ and in  {\bf (d)}  $\tau_{t}^{min}$. See text for details}
\label{fig6}
\end{figure*}

\begin{figure*}[!htb]
\includegraphics[scale=0.45]{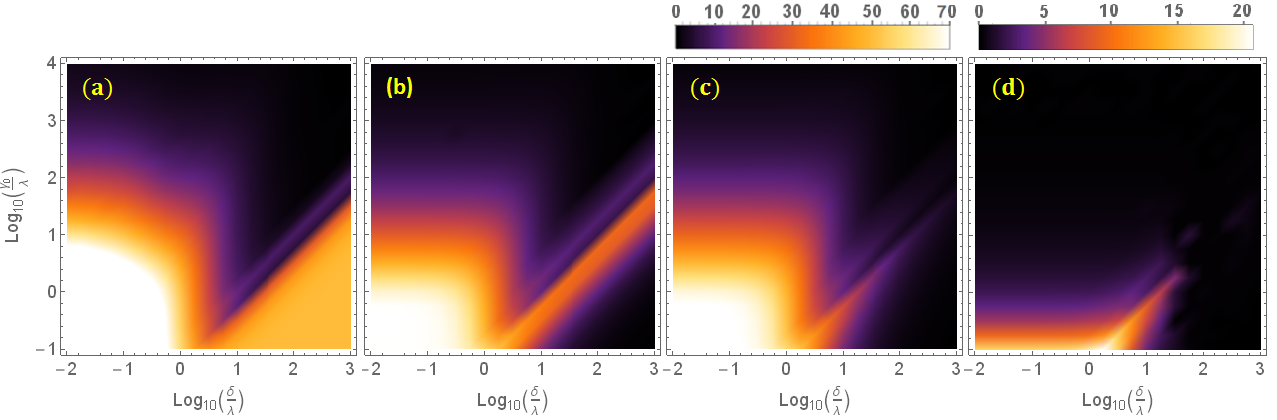}
\caption{(Color online) Density plot of the QSL bounds as a function of the  parameters $\gamma_0/\lambda$ and $\delta/\lambda$.
The initial state is $\hat\rho_{i}=\dyad{x;+}{x;+}$ and the final time of evolution is $\lambda t=100$. In {\bf (a)} we plot  $\tau_{t}^{quant}$,
in  {\bf (b)} $\tau_{t}^{op}$, in {\bf (c)} $\tau_{t}^{av}$ and in  {\bf (d)}  $\tau_{t}^{min}$. 
The density color scale in the range $0-70$ is for pannels 
{\bf (a)}, {\bf (b)} and {\bf (c)}, while the one in the range $0-20$ is for panel {\bf (d)}.  
This shows that $\tau_{t}^{min}$ is a better estimate of the minimal evolution time
between the states $\hat\rho_{i}$ and $\hat \rho_{\lambda t=100}$.}
\label{fig7}
\end{figure*}

In order to compare the non-Markovianity measure $\tilde{\mathcal{N}}$ with the QSL bounds we compute them for the same region of 
parameters
 $\gamma_0/\lambda$ and $\delta/\lambda$ and also considering the initial state $\hat \rho_i=\dyad{x;+}{x;+}$. In Fig. \ref{fig6} we show the QSL bounds 
calculated for a final state at $\lambda t=1$  and in Fig. \ref{fig7} for a final state at $\lambda t=100$. The region of  large $\tilde{\mathcal{N}}$  in Fig.\ref{fig5} {\bf (a)}  corresponds to small values of all 
 of the QSL bounds in Fig. \ref{fig6}. Same result can be observed comparing Fig. \ref{fig5} (b) and  Fig. \ref{fig7}. In this sense,
 large non-Markovianity implies small QSL bounds. 
 This is a manifestation of the speed up of the quantum evolution in the non-Markovian regime
 that we have shown in Fig. \ref{fig2}.  
 But looking at the value of the QSL bounds for different values of the parameters   
 $\gamma_0/\lambda$ and $\delta/\lambda$, it is
not possible to infer which are the parameter regions of non-Markovian behaviour of the channel.
For example, the region of small values of the QSL bounds
in the lower right corner in panels {\bf (b)}, {\bf (c)}, {\bf (d)},  and intermediate values 
in panel {\bf (a)},  of the Fig.\ref{fig6}, do not correspond to the region of parameters with high values of the measure
$\tilde{\mathcal{N}}$ in Fig. \ref{fig5} {\bf (a)} . Exactly the same analysis can be done for the case of $\lambda t=100$ that are plotted in
panel {\bf (b)} of Fig.\ref{fig5} and   Fig.\ref{fig7}.

\section{\label{Section-conclusion}Conclusions}

Two quantum states are not perfect distinguishable unless their supports 
do not overlap. This makes that states that are close in Hilbert space 
are less distinguishable, so the distance between states fix 
the degree of distinguishability between them.
Therefore, in order to connect with a physical evolution two states with some
fix degree of distinguishability, it is necessary to at least go the same distance that separates the two states. This is the origin of the minimal time of evolution settled 
by quantum mechanics.  The Quantum Speed Limit theory is devoted to 
establish lower bounds of this minimal time of evolution
and its origin dates back to the pioneers works of Mandelstam \& Tamm and Margolus \& Levitin
for unitary evolutions connecting pure states. 
It is important to note that the lost of 
the distinguishability between near neighbours states in quantum mechanics is 
intrinsic and has nothing to do with the precision of the measurement apparatus  
used to distinguish them. This contrasts with the classical case where the states of the 
system are given by points in the phase space, whose distinguishability is not related
with the distance between them.  
\par 
A reasonable requirement that any expression corresponding to a QSL bound for the minimal time of evolution between two states
must satisfy is that if we apply the formula in the context of a given dynamics, the result 
must be close to the minimal time of evolution and not to the actual time of evolution
between the states (unless the bound has been saturated).
In this work we analysed the QSL bounds for the minimal time of evolution  in open
quantum systems \cite{taddei,deffner,jing}, and have shown 
that only one, given in \cite{taddei}, effectively verify this basic requirement. 
This was done using the damped Jaynes-Cummings model that for any initial state 
has the same stationary state. So, we have revealed that the QSL bounds in 
 \cite{deffner,jing} grow indefinitely  with the actual evolution time 
 while the final state is essentially reached for finite times. On the contrary 
 the QSL bound in  \cite{taddei} remains constant for any time greater that the time
 where the stationary state is essentially reached. 
 We have also demonstrate that, contrary to the QSL bounds in \cite{taddei, jing}, the QSL bound in  \cite{deffner} can not be saturated continuously in time along a quantum evolution path.
 \par
In relation with the possible link between the non-Markovian effects and 
the behaviour of QSL bounds we found that all of the analysed bounds have lower 
values in a parameter region that match 
the parameter region where takes place 
the speed up of the quantum evolution due to non-Markovian effects in the damped Jaynes-Cummings model. However, we also have shown that there is a 
parameter region of lower values of all the analysed bounds that does not correspond 
to the region of non-Markovian effects on the evolution. In this sense, we
have demonstrated,  with a counterexample, that the statement that the non-Markovian effects on a quantum evolution can be study through the QSL bounds is 
false.

\begin{acknowledgements}
FT acknowledge financial support from the Brazilian agencies CNPq,  CAPES and the INCT-Informa\c{c}\~ao Qu\^antica. 
DAW acknowledge support from CONICET, UBACyT, and ANPCyT (Argentina).
We are grateful to R. L. de Matos Filho, M. M. Taddei, C. Pineda and D. Davalos for fruitful discussions.
\end{acknowledgements}

\end{document}